%% file: main.tex
\DocumentMetadata{}
\documentclass[sigconf, screen, dvipsnames, nonacm]{acmart}

\usepackage{hyperref}
\usepackage[noend]{algorithmic}
\usepackage{algorithm}
\usepackage{multirow}
\usepackage{amsmath}
\usepackage{bbm}
\usepackage{booktabs}
\usepackage[inline]{enumitem} 
\usepackage{subcaption}
\usepackage{bm} 
\usepackage{xcolor}

\usepackage{lipsum}

\usepackage{wrapfig}

\usepackage{arydshln}
\makeatletter
\def\adl@drawiv#1#2#3{
        \hskip.5\tabcolsep
        \xleaders#3{#2.5\@tempdimb #1{1}#2.5\@tempdimb}%
                #2\z@ plus1fil minus1fil\relax
        \hskip.5\tabcolsep}
\newcommand{\cdashlinelr}[1]{%
  \noalign{\vskip\aboverulesep
          \global\let\@dashdrawstore\adl@draw
          \global\let\ adl@draw\adl@drawiv}
  \cdashline{#1}
  \noalign{\global\let\adl@draw\@dashdrawstore
          \vskip\belowrulesep}}
\makeatother

\usepackage{tikz}
\usetikzlibrary{automata, arrows, bayesnet, bending}

\usepackage{tikzscale}

\copyrightyear{2024}
\acmYear{2024}
\setcopyright{none}
\acmConference[RecSys '24]{Proceedings of the 18th ACM Conference on Recommender Systems}{October 14--18, 2024}{Bari, Italy}
\acmBooktitle{Proceedings of the 18th ACM Conference on Recommender Systems (RecSys '24), October 14--18, 2024, Bari, Italy}

\begin{document}
\title{A Simple Model to Estimate Sharing Effects in Social Networks}

\author{Olivier Jeunen}
\affiliation{
  \institution{ShareChat}
  \city{Edinburgh}
  \country{United Kingdom}
}

\begin{abstract}
Randomised Controlled Trials (RCTs) are the gold standard for estimating treatment effects across many fields of science.
Technology companies have adopted A/B-testing methods as a modern RCT counterpart, where end-users are randomly assigned various system variants and user behaviour is tracked continuously.
The objective is then to estimate the causal effect that the treatment variant would have on certain metrics of interest to the business.

When the outcomes for randomisation units ---end-users in this case--- are not statistically independent, this obfuscates identifiability of treatment effects, and harms decision-makers' observability of the system.
Social networks exemplify this, as they are designed to \emph{promote} inter-user interactions.
This interference by design notoriously complicates measurement of, e.g., the effects of sharing.
In this work, we propose a simple Markov Decision Process (MDP)-based model describing user sharing behaviour in social networks. 
We derive an unbiased estimator for treatment effects under this model, and demonstrate through reproducible synthetic experiments that it outperforms existing methods by a significant margin.
\end{abstract}

\maketitle

\input{1_Intro}
\input{2_Method}
\input{3_Experiments}
\input{4_Conclusions}

\bibliographystyle{ACM-Reference-Format}
\bibliography{bibliography}

\end{document}

%% file: 1_Intro.tex
\section{Introduction \& Motivation}
Randomised experiments are the cornerstone of treatment effect estimation~\cite{Fisher1921,Rubin1974}, and modernised extensions of classical methods permeate daily practice in technology companies~\cite{kohavi2020trustworthy}.
Notwithstanding their popularity, the interpretation of A/B-outcomes is error-prone when the assumptions underlying the statistical methodology are either poorly understood or violated~\cite{Kohavi2022, Dmitriev2017,Jeunen2023_Forum}.
One such assumption is the Stable Unit Treatment Value Assumption (SUTVA), which states that the outcome for a unit should be independent of the outcomes of other units who were assigned different variants.

Common violations of the SUTVA occur in the presence of multi-sided marketplaces~\cite{Liu2021,Bajari2021,Farias2023}, machine learning pipelines~\cite{Jeunen2023_Forum}, or so-called ``network effects''~\cite{Gupta2019}.
A common solution for the latter is to adopt cluster-randomised designs~\cite{Hayes2017,Karrer2021}.
There is, nevertheless, a non-trivial engineering cost associated with setting up the backbone that allows for such experiments to run, and even then, the obtained effect size estimates rely heavily on the clusters.

\citeauthor{Farias2022} cast the problem of treatment effect estimation under interference as a policy evaluation problem~\cite{Farias2022}, leveraging reinforcement learning theory to obtain an estimator with a favourable bias-variance trade-off.
Their ``Differences-in-Qs'' estimator has been used to correct for experiment interference at Douyin~\cite{Farias2023}.

We focus on estimating the effects of sharing on consumption metrics: ``\textit{how many additional sessions do we observe due to users sharing content?}''
As long-term business goals are typically centred around growth, the ability to accurately estimate this quantity from user-randomised experiments is crucial.
In this work, we propose a simple Markov Decision Process (MDP)-based model to describe user sharing behaviour.
We derive an unbiased estimator for sharing effects under this model, and empirically demonstrate its efficacy.

%% file: 2_Method.tex
\begin{figure}
\includegraphics{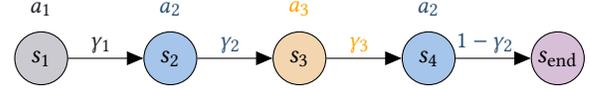}
    \caption{An example trajectory from our MDP: session $s_{1}$ which was assigned system variant $a_{1}$ leads to session $s_{2}$ (variant $a_{2}$), which leads to $s_{3}$ (variant $a_{3}$), and finally $s_{4}$ (variant $a_{2}$). We wish to estimate the expectation of trajectory lengths under constant actions (i.e. shipping a variant to all users).}
    \label{fig:MDP}
\end{figure}
\section{Problem Statement}
Following recent work~\cite{Farias2023}, we adopt an MDP to model user behaviour, consisting of: states $\mathcal{S}$, actions $\mathcal{A}$, transition probabilities $\mathsf{P}_{a}$, and rewards $\mathsf{R}_{a}$.
Actions correspond to system variants being tested in an experiment.
Users are assigned a variant, and they can either share content with another user (starting a Markov chain), or not.
The probability with which they do this ---the transition probability--- depends on the state and sequence of actions taken in the preceding chain thus far.
For simplicity, but w.l.o.g., we model constant $1$ rewards for every ``successful share'' transition, and $0$ reward when a so-called ``sharing chain'' stops.
Note that this can easily be extended to incorporate session-quality metrics instead.
A policy $\pi: \mathcal{S} \to \mathcal{A}$ maps states (users) to actions (system variants).
We typically obtain data from the production policy $\pi_{p}$, stochastically mapping users to variants in an online experiment.
Other policies of interest relate to scaling variants to the population, thus deterministically yielding: $\pi_{a}(A=a|S=s)=1, \forall s\in\mathcal{S}, \forall a \in \mathcal{A}$.

The estimand of interest is the total reward a policy incurs: $V(\pi) = \sum_{t=0}^{\infty}\mathbb{E}_{a\sim \pi}[r_{t}]$.
In online experiments, we typically care about estimating treatment effects: $V_{\Delta}(\pi^{\prime},\pi) \equiv V(\pi^{\prime})-V(\pi)$~\cite{Jeunen2024_DeltaOPE}.

A \textbf{Na\"ive} estimator ignores interference effects, directly estimating $V(\pi)$ via inverse propensity score weighting and subtracting the estimates.
Naturally, this estimator is biased under interference:
\begin{equation}
\widehat{V}_{\Delta}^{N}(\pi_{a_{i}},\pi_{a_{j}}) = \sum_{t=0}^{\infty}\left(\frac{\mathbf{1}(a_{t}=a_{i})}{\pi_{p}(a_{i})}r_{t} - \frac{\mathbf{1}(a_{t}=a_{j})}{\pi_{p}(a_{j})}r_{t}\right).
\end{equation}

The \textbf{Differences-in-Qs} estimator improves on $\widehat{V}_{\Delta}^{N}$ by limiting its myopia~\cite{Farias2023}, replacing observed rewards with Q-value estimates:
\begin{gather}
\widehat{V}_{\Delta}^{Q}(\pi_{a_{i}},\pi_{a_{j}}) \nonumber \\
= \sum_{t=0}^{\infty}\left(\frac{\mathbf{1}(a_{t}=a_{i})}{\pi_{p}(a_{i})}\left(\sum_{t^{\prime}=t}^{\infty}r_{t^{\prime}}\right) - \frac{\mathbf{1}(a_{t}=a_{j})}{\pi_{p}(a_{j})}\left(\sum_{t^{\prime}=t}^{\infty}r_{t^{\prime}}\right)\right).
\end{gather}
This estimator is state-agnostic and exhibits a favourable bias-variance trade-off, making it an attractive method for general use-cases where interference corrections are needed.

\begin{figure*}[!t]
    \centering
    \vspace{-4ex}
    \includegraphics[width=\linewidth]{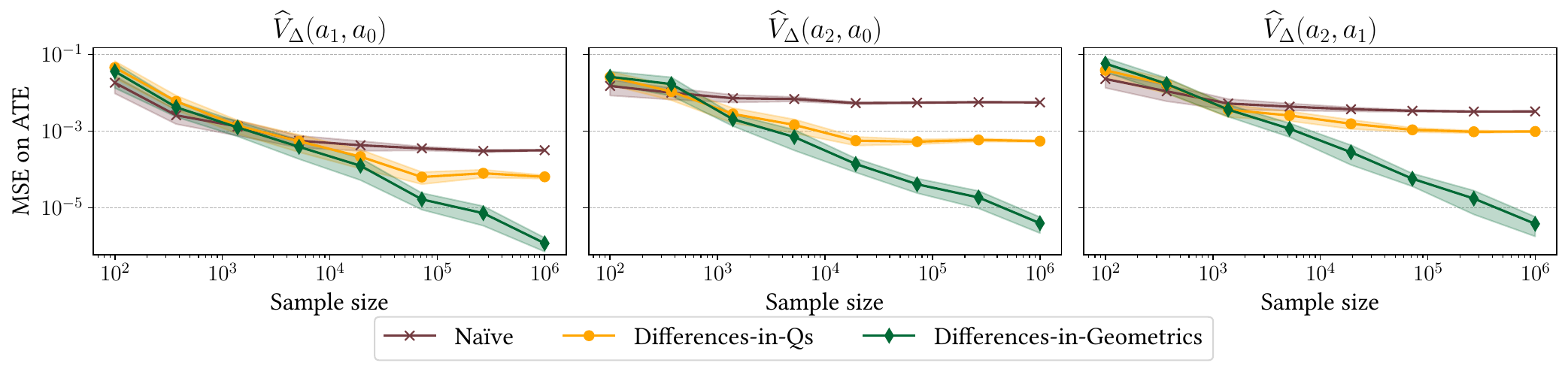}
    \caption{Treatment effect estimation errors for a synthetic setup simulating sharing effects, showing 95\% confidence intervals over 32 repeated runs. We observe that the Differences-in-Geometrics estimator performs favourably compared to alternatives.}\label{fig:exp}
\end{figure*}
\section{Methodology \& Contributions}
Sharing behaviour in social networks encodes a specific type of MDP.
Indeed, one might reasonably assume that whilst the likelihood of a user sharing content is influenced by the system variant they have been assigned, it is independent of what variants were assigned to other users.
As such, in any given state $s$, the probability that the trajectory or ``sharing chain'' ends is given by:
\begin{equation}\label{eq:assm}
    \mathsf{P}_{a}(s_{\rm end}|s) = 1-\gamma_{a}, \qquad \forall s \in \mathcal{S}_{\setminus s_{\rm end}}.
\end{equation}
Empirical estimates for $\gamma_{a}$ can easily be computed from Monte Carlo samples obtained under the production policy.
That is, given a dataset $\mathcal{D}$ of user sessions, their assigned variants ($a$), and whether they led to additional sessions through sharing ($r$), we have:
\begin{equation}
    \widehat{\gamma_{a_{i}}} = \frac{1}{|\mathcal{D}|}\sum_{(a,r) \in \mathcal{D}}\frac{\mathbf{1}(a=a_{i})}{\pi_{p}(a_{i})}r.
\end{equation}
Figure~\ref{fig:MDP} visualises an example trajectory under such an MDP.

Under this assumption, we can rewrite the value of a policy as the expected number of ``successful shares'' before the chain ends:
\begin{equation}\label{eq:geometric}
V(\pi_{a}) = \sum_{t=0}^{\infty}\mathbb{E}_{\pi_{a}}[r_{t}] = \sum_{k=0}^{\infty} k \gamma_{a}^{k-1}(1-\gamma_{a}).
\end{equation}

The infinite sum in the right-hand side of Eq.~\ref{eq:geometric} involves the first derivative of a geometric series, which allows us to obtain a closed-form solution for it.
Recall the geometric sum is given by:
\begin{equation}
    \sum_{k=0}^{\infty} \gamma^k = \frac{1}{1 - \gamma}, \quad \text{for} \quad |\gamma| < 1.    \nonumber 
\end{equation}
Considering its derivative w.r.t. $\gamma$, we obtain:
\begin{equation}
\frac{\mathrm{d}}{\mathrm{d}\gamma} \left( \sum_{k=0}^{\infty} \gamma^k \right) = \frac{\mathrm{d}}{\mathrm{d}\gamma} \left( \frac{1}{1 - \gamma} \right),
\qquad\qquad
\sum_{k=1}^{\infty} k \gamma^{k-1} = \frac{1}{(1 - \gamma)^2}. \nonumber
\end{equation}
As such, we can rewrite the value of a policy as:
\begin{equation}
    V(\pi_{a}) = (1-\gamma_{a})\sum_{k=0}^{\infty} k \gamma_{a}^{k-1} = \frac{1}{1-\gamma_{a}}. 
\end{equation}
This gives rise to the \textbf{Differences-in-Geometrics} estimator:
\begin{equation}
\widehat{V}_{\Delta}^{G}(\pi_{a_{i}},\pi_{a_{j}}) = \frac{1}{1-\widehat{\gamma_{a_{i}}}}-\frac{1}{1-\widehat{\gamma_{a_{j}}}}
\end{equation}
If Eq.~\ref{eq:assm} holds, $\widehat{V}_{\Delta}^{G}$ provides an unbiased estimator of the treatment effect.
Note that, like $\widehat{V}_{\Delta}^{Q}$, our proposed estimator is state-agnostic. 

%% file: 3_Experiments.tex
\section{Empirical Results \& Discussion}
We wish to empirically assess the performance of available treatment effect estimators for the estimation task at hand.
Offline datasets are limited in what they can reveal, and none are publicly available.
For these reasons we resort to a simulated setup, which has the advantage of reproducibility.
All code to reproduce results is available at \textcolor{Maroon}{\href{https://github.com/olivierjeunen/CONSEQUENCES24-sharing}{github.com/olivierjeunen/CONSEQUENCES24-sharing}}.

We simulate three system variants, with $\pi_{p}(A)\coloneq [0.5, 0.25, 0.25]$ and $\bm{\gamma} \coloneq [0.1, 0.2, 0.3]$.
We sample $N$ trajectories, representing chains of users who are assigned various system variants, independently either sharing content to prolong the trajectory or ending it.
This data is used to estimate Average Treatment Effects (ATEs) $V_{\Delta}(\cdot,\cdot)$ for all pairwise policy comparisons.
Figure~\ref{fig:exp} visualises a 95\% confidence interval around the mean squared error on the ATE as we increase the sample size, for all competing estimators.
Empirical observations corroborate our theoretical expectations.
Whilst the Na\"ive estimator suffers from bias due to interference effects, the Differences-in-Qs estimator is able to significantly reduce this.
It is, nevertheless, unable to fully alleviate the bias.
Our proposed estimator that leverages the MDP structure to reframe policy value as a geometric sum is unbiased and consistent in this simulated scenario, outperforming existing approaches by a significant margin.

These results are preliminary, and so far do not provide insights about situations where the independence assumption in Eq.~\ref{eq:assm} is violated.
Nevertheless, they highlight that the performance of general-purpose estimators be improved significantly by making appropriate assumptions about the nature of the MDP and the underlying interference that leads to the SUTVA being violated.

%% file: 4_Conclusions.tex
\section{Conclusions \& Outlook}
Social networks are designed to promote content sharing, as it provides a lever for organic growth of the platform and underlying business. 
As such, when system changes impact sharing effects, it is of paramount importance to measure this accurately.
Classical A/B-testing methodology suffers from interference in such situations, as users can share content with users assigned conflicting variants, or not even in the experiment.
This obfuscates identifiability of treatment effects, inhibiting reliable decision-making as a result.

In this work, we model user sharing behaviour in social networks as an MDP, and cast the problem of estimating sharing effects as one of policy evaluation.
We propose a novel estimator that is unbiased under a mild independence assumption, and highlight its potential via reproducible synthetic experiments.